\documentclass{appolb}
\usepackage{amssymb}
\usepackage[dvips]{graphicx}

\newcommand{\sq}{\sqrt{1 - \frac{2m}{r}}}
\newcommand{\ebind}{E_\mathrm{bind}}
\newtheorem{theorem}{Theorem}
\newtheorem{corollary}{Corollary}
\newtheorem{lemma}{Lemma}

\begin{document}
\pagestyle{plain}
\newcount\eLiNe\eLiNe=\inputlineno\advance\eLiNe by -1
\title{From polytropic to barotropic perfect fluids\\in general-relativistic hydrodynamics}
\author{Bogusz Kinasiewicz and Patryk Mach\\%
\address{M. Smoluchowski Institute of Physics, Jagellonian University,\\Reymonta 4, 30-059 Krak\'{o}w, Poland}}
\maketitle

\begin{abstract}
Two general-relativistic hydrodynamical models are considered: a model of self-gravitating static configurations of perfect fluid and a model of steady accretion of fluid onto a black hole. We generalise analytic results obtained for the original polytropic versions of these models onto a wider class of barotropic equations of state. The knowledge about the polytropic solutions is used to establish bounds on certain characteristic quantities appearing in both cases.
\end{abstract}

\section{Introduction}

We will deal with two general-relativistic hydrodynamical models assuming the spherical symmetry: a model of self-gravitating static configurations of barotropic perfect fluids and a model of steady accretion of barotropic fluid onto a black hole. 

Although the first issue belongs to a classical astrophysics (see e.g. \cite{buchdahl,tooper}), quite recently Karkowski and Malec gave analytic estimates onto the binding energy of the spherically symmetric, compact polytropes \cite{KM}. These estimates had been used to infer the connection between the maximal sound speed and the sign of the binding energy related to the stability of the configuration.

The Newtonian version of the second model \cite{bondi} has been promoted to a general-relativistic one by Michel \cite{michel} within the so-called test fluid approximation. The first fully general-relativistic description taking into account the backreaction (i.e., the impact the accreting fluid has on the space-time geometry) has been formulated by Malec \cite{malec1999}.

Both Karkowski and Malec \cite{KM} and the authors of the works on the steady accretion cited above deal with the polytropic fluids, i.e., fluids with the equation of state expressed either as $\tilde p = K n^\Gamma$ or in the form $\tilde p = K \varrho^\Gamma$, where $K$ and $\Gamma$ are constant. Here $\tilde p$ denotes the isotropic pressure, $n$ is a baryonic density, and $\varrho$ refers to the energy density appearing in the energy-momentum tensor of perfect fluid.\footnote{We will refer to these two kinds of polytropes (coinciding in the Newtonian regime) as to a $\tilde p$ -- $n$ and a $\tilde p$ -- $\varrho$ polytrope respectively.}

In the present paper we show generalisations of these models obtained by considering barotropic equations of state, instead of the polytropes. More precisely: it follows that the knowledge of some properties of the polytropic solutions can be used to estimate those properties in a more general, barotropic case.

Allowing for a wider class of equations of state constitutes a step toward more physical cases than just polytropes. A simple barotropic model of a white dwarf can serve as an example. The simplest white dwarf model is that described by a two component barotropic equation of state with the two components being essentially of the polytropic form: a polytrope with the exponent equal 5/3 for low values of the density and 4/3 in the ultra-relativistic regime.

The order of this paper is as follows. In the next section we give basic definitions of the quantities used in the presented analysis. Section \ref{binding_energy_of_the_static_perfect_fluids} contains the discussion of the first model --- that of static configurations of perfect fluids, where we deal with the estimates onto the binding energy in the barotropic case. Section \ref{quasistationary_accretion} is devoted to the model of the steady accretion of barotropic fluids onto a black hole. At the end some closing remarks are made in section \ref{closing_remarks}.

%%%%%%%%%%%%%%%%%%%%%%%%%%%%%%%%%%%%%%%%%%%%%%%%%%%%%%%%%%%%%%%%%%%%%%%%%%%%%%%
%%%%%%%%%%%%%%%%%%%%%%%%%%%%%%%%%%%%%%%%%%%%%%%%%%%%%%%%%%%%%%%%%%%%%%%%%%%%%%%

\section{Basic equations and definitions}
\label{basic_equations_and_definitions}

In the present section we recollect some of the equations and definitions introduced in \cite{malec1999} and \cite{KM}.

A general, spherically symmetric space-time can be described by a line element
\begin{equation}
\label{line_element}
ds^2 = -N^2 dt^2 + \alpha dr^2 + R^2 \left( d \theta^2 + \sin^2 d \phi^2 \right),
\end{equation}
where $N$, $\alpha$ and $R$ are functions of a coordinate radius $r$ and an asymptotic time variable $t$.

We will consider a foliation of the space-time by the slices of a fixed time $t$. The only non-vanishing elements of the second fundamental form of such slices can be computed to yield $K_r^r = \partial_t \alpha / (2N\alpha)$, $K_\theta^\theta = K_\phi^\phi = \partial_t R / (NR)$
(the convention for defining the second fundamental form used here is that of Wald \cite{wald}). The trace of the second fundamental form reads $\mathrm{tr} K = N^{-1} \partial_t \ln \left( \sqrt{\alpha} R^2 \right)$.
The analogous calculations can be also performed for the two-spheres of constant radius $r$ embedded in each of the time slices. The expression for the trace of the second fundamental form of such sphere is given by $p = 2 \partial_r R / \left( R \sqrt{\alpha} \right)$.

We assume further the energy-momentum tensor of perfect fluid
\begin{equation}
\label{energy-momentum_tensor}
T^{\mu\nu} = (\tilde p + \varrho) u^\mu u^\nu + \tilde p g^{\mu \nu},
\end{equation}
where $u^\mu$ denotes the four velocity of the fluid, $\tilde p$ is the pressure and $\varrho$ the energy density.

The line element of the form (\ref{line_element}) still has certain gauge freedom. While when dealing with static configurations one usually uses the so-called polar gauge, i.e., one characterised by a condition $\mathrm{tr} K = K_r^r$, for the description of the evolving fluids it proves to be convenient to introduce the comoving coordinates. This can be done by imposing an integral condition of the form
\begin{eqnarray*}
\lefteqn{\left( \frac{1}{2} R \left( \mathrm{tr} K - K_r^r \right) \right)^2 = \frac{3}{4R} \int_R^\infty {R^\prime}^2 (K_r^r)^2 dR^\prime +} \\
& & - \frac{1}{2R} \int_R^\infty {R^\prime}^2 K_r^r \mathrm{tr} K dR^\prime - \frac{1}{4R} \int_R^\infty {R^\prime}^2 \mathrm{tr} K dR^\prime.
\end{eqnarray*}
One can check that this indeed gives $u^r = u^\theta = u^\phi = 0$ \cite{malec1999}.

The relation $\nabla_\mu T^{\mu \nu} = 0$ leads to the continuity equation
\begin{equation}
\label{continuity_eqn}
\nabla_\mu \left( \varrho u^\mu \right) = - \tilde p \nabla_\mu u^\mu.
\end{equation}
Given a barotropic equation of state of the form $\tilde p = \tilde p(\varrho)$ it is, however, possible to find an integrability factor $n$ (the so-called baryonic density) such that $\nabla_\mu \left( n u^\mu \right) = 0$. It can be shown that for the energy-momentum tensor (\ref{energy-momentum_tensor}) the function
\begin{equation}
\label{n_as_a_function_of_rho}
n = n_1 \exp \int_{\varrho_1}^\varrho \frac{d \varrho^\prime }{\varrho^\prime + \tilde p(\varrho^\prime) }
\end{equation}
has the desired property. In turn, if we assume an equation of state expressed as $\tilde p = \tilde p(n)$ then the function $\varrho$ given by
\begin{equation}
\label{rho_as_a_function_of_n}
\varrho = n + n \int_0^n dn^\prime \frac{\tilde p (n^\prime) }{{n^\prime}^2}
\end{equation}
satisfies the continuity equation (\ref{continuity_eqn}). Equations (\ref{n_as_a_function_of_rho}) and (\ref{rho_as_a_function_of_n}) can be treated as expressing the relations between the baryonic and the energy density.

For a given baryonic density $n$ one defines the baryonic mass by a simple integration over the volume of the configuration. The baryonic mass enclosed in a ball $B(r)$ of the coordinate radius $r$ is thus equal to $m_\mathrm{bar}(r) = \int_{B(r)} dV n = 8 \pi \int_0^{R(r)} dR^\prime \left( R^\prime n / p \right)$. Here $dV$ denotes the volume element induced on the hypersurfaces of constant time $t$.

An analogous volume integral of the energy density defines the so-called rest mass $M(r) = \int_{B(r)} dV \varrho = 8 \pi \int_0^{R(r)} dR^\prime \left( R^\prime \varrho / p \right)$.
Although it is clearly a geometric quantity on a fixed Cauchy hypersurface, it is not conserved during the evolution of the system, i.e., it can change between subsequent Cauchy time slices.

A mass measure that, at least for compact systems, can be shown to be conserved (see e.g. \cite{malec1999}), and that is defined in terms of the energy density is the quasilocal mass $m(r) = \int_{B(r)} dV \left( Rp\varrho / 2 \right) = 4 \pi \int_0^{R(r)} dR^\prime {R^\prime}^2 \varrho$.

The quasilocal mass of a static (and spherically symmetric) configuration of perfect fluid enclosed entirely within a sphere of an areal radius $R$ coincides with the asymptotic mass. It will be denoted here as $m_S = m(r(R)) = 4 \pi \int_0^R dR^\prime {R^\prime}^2 \varrho$. Similarly to $m_S$, the symbols $M_S$ and $m_\mathrm{bar}$ will refer to the total rest mass and the total baryonic mass of the static, spherically symmetric system contained in a sphere of an areal radius $R$.

Let us also note that one can write down the line element for the static spacetime in the polar gauge as
\[ ds^2 = -N^2 dt^2 + \frac{1}{1 - \frac{2m}{r}} dr^2 + r^2 \left( d\theta^2 + \sin^2 d \phi^2 \right). \]
In this case both $N$ and $m$ are functions of the (areal) radius $r$.

%%%%%%%%%%%%%%%%%%%%%%%%%%%%%%%%%%%%%%%%%%%%%%%%%%%%%%%%%%%%%%%%%%%%%%%%%%%%%%%
%%%%%%%%%%%%%%%%%%%%%%%%%%%%%%%%%%%%%%%%%%%%%%%%%%%%%%%%%%%%%%%%%%%%%%%%%%%%%%%

\section{Binding energy of the static perfect fluids}
\label{binding_energy_of_the_static_perfect_fluids}

The binding energy of the static, spherically symmetric and compact configuration of the perfect fluid is usually defined as a difference between its baryonic and asymptotic mass, i.e.,
$\ebind = m_\mathrm{bar} - m_S = \int_V dV n - \int_V dV \varrho \sqrt{1 - 2m/r}$.
The paper by Karkowski and Malec \cite{KM} is mainly devoted to give an analytic estimate of this energy for the polytropic configuration of fluid. However, one of the most useful results obtained there is a theorem bounding the potential energy $-\tilde U_G$ defined as
\begin{eqnarray*}
-\tilde U_G & = & M_S - m_S = \int_V dV \varrho \left( 1 - \sq \right) = \\
& = & \int_V dV \frac{2 \varrho m}{r \left( 1 + \sq \right)}.
\end{eqnarray*}

\begin{theorem}[Karkowski, Malec]
\label{th_karkowski_malec}
Consider a static spherically symmetric configuration of the self-gravitating fluid contained entirely within a sphere of the areal radius $R$. Let $\varrho \geqslant 0$. Then
\begin{eqnarray*}
\lefteqn{- \tilde U_G = M_S - m_S  \geqslant} \\
& \geqslant  & \frac{m_S^2}{R}\frac{2}{\left( 1 + \sqrt{1 - \frac{2 m_S}{R}} \right)^2} - \frac{m_S}{2\left( 1 + \sqrt{1 - \frac{2m_S}{R}} \right)} + \\
& & + \frac{3R}{4} \left( -1 + \sqrt{\frac{R}{2m_S}} \arcsin \sqrt{\frac{2m_S}{R}} \right) \geqslant \frac{3 m_S^2}{5R}.
\end{eqnarray*}
\end{theorem}

Note that this is a very general statement --- no explicit form of the equation of state is being assumed.

For the sake of shortness we will denote by $\Xi$ the right-hand side of the above inequality.

As an implication of this theorem one can write the following bound on the binding energy of the static, polytropic configuration
\begin{equation}
\label{polytropic_bound_on_the_binding_energy}
E_\mathrm{bind} \geqslant M_S - \frac{P}{\Gamma - 1} - m_S \geqslant \Xi - \frac{P}{\Gamma - 1},
\end{equation}
where $P$ denotes the integral $P = \int_V dV \tilde p$. Below we will give the inequalities of this kind, valid for a wider class of barotropic equations of state.

%%%%%%%%%%%%%%%%%%%%%%%%%%%%%%%%%%%%%%%%%%%%%%%%%%%%%%%%%%%%%%%%%%%%%%%%%%%%%%%

\subsection{Binding energy of the barotropic fluid spheres}
\label{binding_energy_of_the_barotropic_fluid_spheres}

In many cases considered in astrophysics barotropic equations of state do not differ much from the polytropes. This suggests writing a barotropic equation of state as
\begin{equation}
\label{barotropic_eos_n}
\tilde p(n) = \tilde p (n_0) \left( \frac{n}{n_0} \right)^{\Gamma(n,n_0)},
\end{equation}
where $n_0$ denotes an arbitrarily chosen density of reference and the exponent $\Gamma$ is a function of $n$.

The function $\Gamma$ introduced in this way appears to be a kind of a mean value of the so-called adiabatic index $\Gamma_1 = \left( n / \tilde p \right) d \tilde p / d n = \left( \left( \varrho + \tilde p \right) / \tilde p \right) d \tilde p / d \varrho $. The two ``gammas'' are related by the formula
\[
\Gamma(n,n_0) = \ln^{-1}  \frac{n}{n_0} \int_{n_0}^n d n^\prime \frac{\Gamma_1(n^\prime)}{n^\prime}.
\]

Clearly, every barotropic equation with $n \geqslant 0$ and $\tilde p \geqslant 0$ can be written in the form (\ref{barotropic_eos_n}). We will, however, deal with a narrower class of equations of state for which the function $\Gamma$ is bounded, i.e., $\Gamma_\mathrm{min} \leqslant \Gamma(n,n_0) \leqslant \Gamma_\mathrm{max}$, and that the inequality $\Gamma_\mathrm{min} > 1$ holds. This implies that $\tilde p(n = 0) = 0$ and also that $\tilde p(n)/n \to 0$ as $n \to 0$. These properties will be used explicitly in further considerations.

We present now two generalisations of (\ref{polytropic_bound_on_the_binding_energy}) onto the case with the equations of state of the form (\ref{barotropic_eos_n}).

\begin{lemma}
\label{th_non_increasing_gamma_n}
Assume conditions of theorem \ref{th_karkowski_malec}. Let the equation of state be of the form (\ref{barotropic_eos_n}) and let the exponent $\Gamma$ be nonincreasing with the baryonic density $n$. If the reference density $n_0$ is such that $n \leqslant n_0$, then $\ebind \geqslant \Xi - P / \left( \Gamma_\mathrm{min} - 1 \right).$
\end{lemma}

Proof. Let us start with the inequality
\[ n \int_0^n \frac{dn^\prime}{n_0} \frac{\tilde p(n_0)}{n_0} \left( \frac{n^\prime}{n_0} \right)^{\Gamma(n,n_0) - 2} \geqslant n \int_0^n \frac{dn^\prime}{n_0} \frac{\tilde p(n_0)}{n_0} \left( \frac{n^\prime}{n_0} \right)^{\Gamma(n^\prime,n_0)} \]
valid for $n \leqslant n_0$ and $\Gamma(n,n_0)$ being a nonincreasing function of $n$. The integral appearing on the left-hand side of the above formula can be easily computed to give
\[ \frac{\tilde p (n_0)}{\Gamma(n,n_0) - 1} \left( \frac{n}{n_0} \right)^{\Gamma (n,n_0)} \geqslant n \int_0^n dn^\prime \frac{\tilde p (n^\prime)}{{n^\prime}^2} \]
and thus $n \geqslant \varrho - \tilde p (n) / (\Gamma(n,n_0) - 1) \geqslant \varrho - \tilde p (n) / \left( \Gamma_\mathrm{min} - 1 \right) $.  It is now enough to use the bound on $- \tilde U$ given in \cite{KM}, i.e., $\int_V dV \varrho - m_S \geqslant \Xi$ to obtain the desired estimate onto the binding energy, namely
\[ \ebind = \int_V dV n - m_S \geqslant \Xi - \frac{1}{\Gamma_\mathrm{min} - 1} \int_V dV \tilde p. \]
This completes the proof.

The similar result can be formulated for the case with $\Gamma(n,n_0)$ being a nondecreasing function of $n$. We decided to omit the proof. Although it is more complex than the preceding one, it can be based on the similar idea.

\begin{lemma}
\label{th_non_decreasing_gamma_n}
Let the assumptions of theorem \ref{th_karkowski_malec} hold and let $\Gamma(n,n_0)$ be a nondecreasing function of $n$. Then
\begin{eqnarray}
\ebind & \geqslant & A \Xi + A m_S - m_S - \frac{A}{\Gamma_\mathrm{min} - 1} \int_V dV \Theta(n - n_0) \tilde p + \nonumber \\
& & - \frac{1}{\Gamma_\mathrm{min}} \int_V dV \Theta(n_0 - n) \tilde p(n_0) \left( \frac{n}{n_0} \right)^{\Gamma_\mathrm{min}},
\label{bound_for_non_decreasing_gamma}
\end{eqnarray}
where $\Theta$ denotes the Heaviside function and the $A$ constant is given by
\[ A = \left( 1 + \frac{\tilde p (n_0)}{n_0}\frac{\Gamma_\mathrm{max} - \Gamma_\mathrm{min}}{\left(\Gamma_\mathrm{max} - 1\right)\left(\Gamma_\mathrm{min} - 1\right)} \right)^{-1}. \]
\end{lemma}

Let us note that in lemma \ref{th_non_decreasing_gamma_n} no assumption is made on the value of the reference density $n_0$.

Consider now the limit of $n_0 \to 0$: (i) Since the volume of the region where $n < n_0$ tends to zero so does the last integral occurring in (\ref{bound_for_non_decreasing_gamma}). (ii) In general by changing $n_0$ we change the form of $\Gamma(n,n_0)$. Suppose now that the expression $\Gamma_\mathrm{max} - \Gamma_\mathrm{min}$ remains bounded as $n_0 \to 0$. Remembering that we are still dealing with the equations of state for which $\tilde p(n)/n \to 0$ for $n \to 0$ we see that the value of $A$ appearing in (\ref{bound_for_non_decreasing_gamma}) tends to the unity. (iii) Assume further that $\Gamma(n,n_0)$ is not only a nondecreasing function for a given $n_0$ but is a nondecreasing function for all $n_0 \to 0$ so that the lemma \ref{th_non_decreasing_gamma_n} holds. Then in the above limit bound (\ref{bound_for_non_decreasing_gamma}) reduces to $\ebind \geqslant \Xi -  P /  \left( \Gamma_\mathrm{min} - 1 \right)$ and the same inequality emerges as a result of both lemmas.

The barotropic equation of state can be also expressed in terms of the energy density $\varrho$. It is clear, that the barotropic equation of state expressed in the form $\tilde p = \tilde p(n)$ can be transformed into the barotropic form in which the energy density is used, i.e., $\tilde p = \tilde p(\varrho)$; in some cases it is, however, useful to preserve the freedom of choice between these two dual descriptions.

In the following we present a result, analogous to the preceding one, but formulated for the equation of state of the form
\begin{equation}
\label{barotropic_eos_rho}
\tilde p (\varrho) = \tilde p (\varrho_0) \left( \frac{\varrho}{\varrho_0} \right)^{\Gamma(\varrho, \varrho_0)}.
\end{equation}
As before we assume that the exponent $\Gamma(\varrho, \varrho_0)$ is bounded, i.e., $1 < \Gamma_\mathrm{min} \leqslant \Gamma(\varrho, \varrho_0) \leqslant \Gamma_\mathrm{max}$ for some constants $\Gamma_\mathrm{min}$ and $\Gamma_\mathrm{max}$.

\begin{lemma}
\label{th_non_increasing_gamma_rho}
Let the conditions of theorem \ref{th_karkowski_malec} hold. Let the equation of state be of the form (\ref{barotropic_eos_rho}) and the exponent $\Gamma(\varrho, \varrho_0)$ be a nonincreasing function of $\varrho$. Let us also assume that the density of reference $\varrho_0$ is greater than the largest value of $\varrho$ within the whole configuration. Then $\ebind \geqslant \Xi - P / \left( \Gamma_\mathrm{min} - 1 \right)$.
\end{lemma}

The last statement presented in this section gives the suitable bound on the binding energy for the equations of state of type (\ref{barotropic_eos_rho}) with the exponents $\Gamma(\varrho, \varrho_0)$ being nondecreasing functions of $\varrho$.

\begin{lemma}
\label{th_non_decreasing_gamma_rho}
Let the assumptions of theorem \ref{th_karkowski_malec} be satisfied. Let the equation of state be of the form (\ref{barotropic_eos_rho}) and let $\Gamma(\varrho, \varrho_0)$ be nondecreasing function of $\varrho$. Under these conditions the following inequality holds
\begin{eqnarray*}
\ebind & \geqslant & \tilde A \Xi + \tilde A m_S - m_S - \frac{\tilde A}{\Gamma_\mathrm{min} - 1} \int_V dV \Theta (\varrho - \varrho_0) \tilde p + \\
& & - \int_V dV \Theta(\varrho_0 - \varrho) \tilde p(\varrho_0) \left( \frac{\varrho}{\varrho_0} \right)^{\Gamma_\mathrm{min}}.
\end{eqnarray*}
Here the constant $\tilde A$ is given by
\[ \tilde A = \left( 1 + \frac{\tilde p(\varrho_0)}{\varrho_0} \right)^{-\frac{\Gamma_\mathrm{max} - \Gamma_\mathrm{min}}{\left(\Gamma_\mathrm{max} - 1\right)\left(\Gamma_\mathrm{min} - 1\right)}}. \]
\end{lemma}

Similarly, under restrictions analogous to those that have been mentioned in the comment to lemma \ref{th_non_decreasing_gamma_n}, the $\varrho_0 \to 0$ limit of the above inequality reads
$\ebind \geqslant \Xi - P / \left( \Gamma_\mathrm{min} - 1 \right)$.

%%%%%%%%%%%%%%%%%%%%%%%%%%%%%%%%%%%%%%%%%%%%%%%%%%%%%%%%%%%%%%%%%%%%%%%%%%%%%%%

\subsection{On the sound velocity and the sign of the binding energy}
\label{on_the_sound_velocity}

The results presented in the previous section allow us to formulate conditions involving the sign of the binding energy that remain in a strict analogy to those given in \cite{KM} for the polytropic equations of state.

Assume that the equation of state is given by (\ref{barotropic_eos_rho}). Then the sound velocity can be easily calculated to yield
\begin{eqnarray*}
a^2 & = & \tilde p \left( \ln \left( \frac{\varrho}{\varrho_0} \right) \frac{\partial \Gamma (\varrho, \varrho_0)}{\partial \varrho} + \frac{\Gamma (\varrho, \varrho_0)}{\varrho} \right) = \\
& = & \tilde p \ln \left( \frac{\varrho}{\varrho_0} \right) \frac{\partial \Gamma (\varrho, \varrho_0)}{\partial \varrho} + \Gamma(\varrho, \varrho_0) \frac{\tilde p(\varrho_0)}{\varrho_0} \left( \frac{\varrho}{\varrho_0} \right)^{\Gamma (\varrho, \varrho_0) - 1}.
\end{eqnarray*}
Let us introduce now the mean sound velocity $\bar a$ as
\[ \bar a^2 = \frac{P}{\int_V dV \left( \ln \left( \frac{\varrho}{\varrho_0} \right) \frac{\partial \Gamma (\varrho, \varrho_0)}{\partial \varrho} + \frac{\Gamma (\varrho, \varrho_0)}{\varrho} \right)^{-1}}. \]
Clearly, the quantity defined in this way is bounded from above and from below by the maximal and minimal sound speed respectively. Let us further define $\tilde a^2 = P \Gamma_\mathrm{min} / M_S$. If the exponent $\Gamma$ is a nonincreasing function of $\varrho$ and the density of reference $\varrho_0$ is chosen to be such that $\varrho \leqslant \varrho_0$ or conversely, if $d\Gamma / d\varrho \geqslant 0$ and $\varrho_0 \leqslant \varrho$, then $ \left( \ln \left( \varrho / \varrho_0 \right) \partial \Gamma (\varrho, \varrho_0) / \partial \varrho + \Gamma (\varrho, \varrho_0) / \varrho \right)^{-1} \leqslant \varrho / \Gamma (\varrho, \varrho_0)$. Consequently $\int_V dV \left( \ln \left( \varrho / \varrho_0 \right) \partial \Gamma (\varrho, \varrho_0) / \partial \varrho + \Gamma (\varrho, \varrho_0) / \varrho \right)^{-1} \leqslant M_S / \Gamma_\mathrm{min}$. This, in turn, implies that $\bar a^2 \geqslant \tilde a^2$.

\begin{corollary}
\label{first_negativity_condition}
Under the assumptions described above, the inequality $\bar a^2 > \Gamma_\mathrm{min} \left( \Gamma_\mathrm{min} - 1 \right) \Xi / M_S$.
is a necessary condition for the binding energy $\ebind$ to be negative.
\end{corollary}

\begin{corollary}
Assume that the conditions specified at the beginning of this section are satisfied. Then for the binding energy to be negative it is necessary that
\begin{equation}
\label{second_negativity_condition}
a_\mathrm{max}^2 \geqslant \Gamma_\mathrm{min} \left( 1 + \frac{\Xi}{m_S} \right)^{\Gamma_\mathrm{min} - 1} - \Gamma_\mathrm{min},
\end{equation}
where $a^2_\mathrm{max}$ denotes the maximum value of the sound speed.
\end{corollary}

A brief discussion of inequalities of the type of inequality (\ref{second_negativity_condition}) can be found in \cite{KM}.

%%%%%%%%%%%%%%%%%%%%%%%%%%%%%%%%%%%%%%%%%%%%%%%%%%%%%%%%%%%%%%%%%%%%%%%%%%%%%%%

\subsection{Physical examples}
\label{a_physical_example}

So far we have only been dealing with fairly general assumptions about the equation of state. It is, however, easy to find explicit physical examples of equations of state that satisfy the assumptions we have made. Let us begin with nonincreasing ``gammas''. Consider a simple model of a white dwarf supported entirely by a degenerate electron pressure (see e.g. \cite{padmanabhan}). The pressure of the noninteracting relativistic electron gas in the zero temperature limit can be described by the formula (we use the cgs units in this section to present the formulae in the most familiar form)
\begin{equation}
\label{degenerate_electrons_eos}
\tilde p = \frac{\pi m_\mathrm{e}^4 c^5}{3 h^3} \left[ x (2x^2 - 3) \sqrt{1 + x^2} + 3 \sinh^{-1} x \right],
\end{equation}
where $m_\mathrm{e}$ denotes the rest mass of the electron, $c$ stands for the speed of light and $h$ is the Planck constant. The parameter $x$ appearing here is related to the number density of electrons $n_\mathrm{e}$ by $x = \left( m_\mathrm{e} c / h \right) \left( 3 n_\mathrm{e} / (8 \pi) \right)^{1/3}$. For the white dwarfs that contain almost no hydrogen one can use the following relation between the number density of the electrons $n_\mathrm{e}$ and the baryonic density: $n_\mathrm{e} = n / \left( 2 m_\mathrm{H} \right)$, where $m_\mathrm{H}$ denotes the rest mass of the hydrogen atom. In this case formula ($\ref{degenerate_electrons_eos}$) defines a barotropic equation of state which can be trivially expressed in the form (\ref{barotropic_eos_n}) by taking $\Gamma(n,n_0) = \left( \ln \tilde p(n) - \ln \tilde p(n_0) \right) / \left( \ln n - \ln n_0 \right)$.
One can show that the function $\Gamma$ is always decreasing; moreover, independently of the value of the density $n_0$ we have $\Gamma(n,n_0) \to 5/3$ for $n \to 0$ and $\Gamma(n,n_0) \to 4/3$ for $n \to \infty$.

Examples of the equation of state with nondecreasing barotropic indices, satisfying conditions of statement \ref{th_non_decreasing_gamma_n}, can be found in \cite{van_riper,romero} or \cite{dimmelmeier}. These equations had been used for the description of the stiffening of matter at supranuclear densities. They are essentially of the form (\ref{barotropic_eos_n}) with the exponent $\Gamma$ given by $\Gamma = \Gamma_\mathrm{min} + \eta \Theta(n - n_\mathrm{nuc}) \ln \left( n / n_\mathrm{nuc} \right)$, where $\Gamma_\mathrm{min}$ and $\eta$ are some constants, $n_\mathrm{nuc}$ denotes the nuclear density and $\Theta$ is the Heaviside function. The density of reference $n_0$ is chosen to be equal unity in the used system of units. Let us note that the function $\Gamma$ defined in this way will still be nondecreasing, even if we allow the reference density $n_0$ to tend to zero. Additionally, if we assume $\Gamma_\mathrm{max} = \Gamma(n_\mathrm{max},n_0)$ for some fixed $n_\mathrm{max}$, then $\Gamma_\mathrm{max} \to \Gamma_\mathrm{min}$ as $n_0 \to 0$. Thus, this kind of equation not only fulfills the assumptions of lemma \ref{th_non_decreasing_gamma_n} but it also assures the desired limit behaviour at $n_0 \to 0$.

%%%%%%%%%%%%%%%%%%%%%%%%%%%%%%%%%%%%%%%%%%%%%%%%%%%%%%%%%%%%%%%%%%%%%%%%%%%%%%%
%%%%%%%%%%%%%%%%%%%%%%%%%%%%%%%%%%%%%%%%%%%%%%%%%%%%%%%%%%%%%%%%%%%%%%%%%%%%%%%

\section{Quasistationary (steady) accretion}
\label{quasistationary_accretion}

We will consider a spherically symmetric cloud of gas falling onto a non-rotating black hole. The black hole provides the simplest choice for the central object as one can assume that no shock waves occur.

In this chapter we will work in comoving coordinates. Let us define a fluid velocity $U$ by $U = \partial_t R / N = R \left( \mathrm{tr} K - K_r^r \right) / 2$. This allows us to write the Einstein equations in the form (see e.g. \cite{KM})
\begin{equation}
\label{hamiltonian_constraint}
pR = 2 \sqrt{1 - \frac{2m(R)}{R} + U^2}
\end{equation}
(the Hamiltonian constraint) and
\begin{equation}
\label{momentum_constraint}
\partial_R \left( R^2 U \right) - R^2 \mathrm{tr} K = 0
\end{equation}
(the momentum constraint). The evolution equation reads
\begin{equation}
\label{evolution_eqn}
\partial_t U = \frac{1}{4}(pR)^2 \partial_R N - \frac{N m(R)}{R^2} - 4 \pi R N \tilde p.
\end{equation}
The continuity equation (\ref{continuity_eqn}) can be now written as
\begin{equation}
\label{comoving_continuity_eqn}
\partial_t \varrho = - N \mathrm{tr} K (\varrho + \tilde p).
\end{equation}
The conservation of the energy-momentum tensor provides us also with the relativistic version of the Euler equation 
\begin{equation}
\label{euler_eqn}
N \partial_R \tilde p + (\varrho + \tilde p) \partial_R N = 0.
\end{equation}

We will assume that the accretion is steady and relatively slow, so that the central mass can be regarded as constant (see e.g. \cite{courant}). More precisely: (i) the accretion rate, defined as $\dot m = ( \partial_t - (\partial_t R)\partial_R)m(R)$ for the given areal radius $R$, is assumed to be constant in time; (ii) the fluid velocity $U$, energy density $\varrho$, sound velocity $a$ etc. should remain constant on the surface of fixed $R$: $(\partial_t - (\partial_t R) \partial_R) X = 0$, where $X = U, \varrho, a, \dots$

The accretion rate introduced above can be computed to yield
\begin{equation}
\label{accretion_rate}
\dot m(R) = - 4 \pi N R^2 U (\varrho + \tilde p).
\end{equation}
It was proved by Malec \cite{malec1999} that under the above conditions the accretion rate is independent of the surface (characterised by a given $R$) for which it is calculated, i.e., $\partial_R \dot m=0$ (cf. equation (\ref{mAB}) below).

Under the above assumptions the original partial differential equations become ordinary ones. The evolution equation (\ref{evolution_eqn}) can be written as $NU \partial_R U = (pR)^2 \partial_R N / 4 - N m(R) / R^2 - 4 \pi N \tilde p$, while differentiation of equation (\ref{hamiltonian_constraint}) gives $NU \partial_R U = N (Rp) \partial_R (Rp) / 4 - N m(R) / R^2 + 4 \pi R N \varrho$. By subtracting these equations we obtain one of the key equations of the presented model, namely
\begin{equation}
\label{lapse_eqn}
\partial_R \ln \left( \frac{N}{pR} \right) = \frac{16 \pi}{p^2 R} (\varrho + \tilde p).
\end{equation}
Finally, under the assumption of the quasistationarity of the accretion, the momentum constraint (\ref{momentum_constraint}) together with the continuity equation yields
\begin{equation}
\label{U_eqn}
\partial_R \ln \left( |U| R^2 \right) = - \frac{\partial_R \varrho}{\varrho + \tilde p}.
\end{equation}
Both of these equations can be integrated starting from the outer boundary of the accreting cloud. If $N_\infty$, $p_\infty$ and $R_\infty$ denote the adequate values at the boundary, we can write
\begin{equation}
\label{integrated_lapse_eqn}
\frac{N}{pR} = \frac{N_\infty}{p_\infty R_\infty} \exp \left( - 16 \pi \int_R^{R_\infty} dR^\prime \frac{\varrho + \tilde p}{p^2 R^\prime} \right).
\end{equation}
Similarly integrating equation (\ref{U_eqn}) we get
\begin{equation}
\label{integrated_U_eqn}
U = \frac{A}{R^2 n},
\end{equation}
where $A$ is an integration constant. Analogous algebraic form can be given also to the Euler equation (\ref{euler_eqn}), namely
\begin{equation}
\label{integrated_euler_eqn}
N = \frac{Bn}{\varrho + \tilde p}.
\end{equation}
Here $B$ denotes another integration constant. One can check the validity of this formula using (\ref{n_as_a_function_of_rho}).

Let us note that formula (\ref{accretion_rate}) reduces to
\begin{equation}
\label{mAB}
\dot m = - 4 \pi A B.
\end{equation}
In principle, one can use $\dot m$ instead of a constant $A$. This is what was done in the classical analysis by Bondi \cite{bondi}; we will, however, leave the constant $A$ as it is, in order to preserve the clarity of the equations.

In what follows we will also put $N_\infty = 1$. This can always be done provided that a suitable foliation of the external Schwarzschild space-time is chosen.

Let us now define a sonic point as such, where the length of the spatial velocity vector equals the speed of sound $|\vec U| = 2 U / (pR) = a$. In the Newtonian limit the above definition coincides with the standard requirement of the equality between the velocity of the fluid and the local sound speed. In the following we will denote by the asterisk all values referring to the sonic point.

Below we show a theorem being a simple generalisation (onto the barotropic fluids case) of Malec's result for the type $\tilde p$ -- $\varrho$ polytrope.

\begin{theorem}
For the barotropic fluid accreting in the quasistationary way, the following equation holds at the sonic point
\begin{equation}
\label{sonic_point_rel}
a_\ast^2 \left( 1 - \frac{3 m(R_\ast)}{2R_\ast} + c_\ast \right) = U_\ast^2 = \frac{m(R_\ast)}{2R_\ast} + c_\ast,
\end{equation}
where $c_\ast$ abbreviates the expression
$c_\ast = 2 \pi R^2 \tilde p_\ast$.
\end{theorem}

Proof. The equation (\ref{lapse_eqn}) can be easily written in the form
\[
\partial_R \ln N = \frac{4}{p^2 R^3} \left( \frac{1}{4} p R^2 \partial_R (Rp) + 4 \pi R^2 \varrho + 4 \pi R^2 \tilde p \right).
\]
Now, differentiation of the squared Hamiltonian constraint equation (\ref{hamiltonian_constraint}) gives
\begin{eqnarray*}
\frac{1}{4} p R^2 \partial_R(Rp) + 4 \pi \varrho R^2 & = & \frac{m(R)}{R} + R U \partial_R U = \\
& = & \frac{m(R)}{R} - 2U^2 + \frac{1}{2R^3} \partial_R \left( U^2 R^4 \right)
\end{eqnarray*}
and thus
\[
\partial_R \ln N = \frac{4}{p^2 R^3} \left( \frac{m(R)}{R} - 2U^2 + \frac{1}{2R^3}\partial_R \left( U^2R^4 \right) + 4 \pi R^2 \tilde p \right).
\]
An other expression for the logarithm of the lapse derivative can obtained from the Euler equation (\ref{euler_eqn}), namely $\partial_R \ln N = - \partial_R \tilde p (\varrho + \tilde p)$. Introducing the sound velocity and remembering equation (\ref{U_eqn}) we get (under assumption that the fluid is ruled by the barotropic equation of state) $\partial_R \left( U^2 R^4 \right) = \left(2 / a^2 \right) \partial_R N$. The above equations lead easily to a key relation
\[
\left( 1 - \frac{4U^2}{p^2 a^2 R^2} \right) \partial_R \left( U^2 R^4 \right) = \frac{16 U^2 R}{a^2 p^2} \left( \frac{m(R)}{2R} - U^2 + 2 \pi R^2 \tilde p  \right).
\]
This equation was first obtained by Malec \cite{malec1999}, but with an explicit use of the polytropic equation of state. The proof can now be completed immediately by writing the above equation together with the equation (\ref{hamiltonian_constraint}) at the sonic point.

%%%%%%%%%%%%%%%%%%%%%%%%%%%%%%%%%%%%%%%%%%%%%%%%%%%%%%%%%%%%%%%%%%%%%%%%%%%%%%%%

\subsection{A case without backreaction}
\label{a_case_without_backreaction}

If the mass of accreting fluid is negligible compared to the total asymptotic mass $m$, i.e.,
\begin{equation}
\label{test_fluid}
4 \pi \int_{R > 2m} dR^\prime {R^\prime}^2 \varrho \ll m,
\end{equation}
equation (\ref{integrated_lapse_eqn}) yields $N \approx pR / 2 \approx \sqrt{1 - 2m / R + U^2}$. Now, under an additional condition $c_\ast = 2 \pi R_\ast^2 \tilde p_\ast \ll 2m / R_\ast $ equation (\ref{sonic_point_rel}) can be rewritten in the form $a_\ast^2 \left( 1 - 3m / \left( 2R_\ast \right) \right) = U_\ast^2 = m / \left( 2R_\ast \right)$. Notice, that due to (\ref{test_fluid}) we have $m(R) \approx m$ for $R > 2m$ and we are, in fact, considering the motion of the infalling fluid in the fixed Schwarzschild space-time (the so-called test fluid approximation).

The steady accretion model of Malec \cite{malec1999} has its continuation in a recent paper by Karkowski et al. \cite{kkmms}. The main aim of this work was to investigate the effects of backreaction of fluid onto the metric. It appeared that, at least in the polytropic case ($\tilde p$ -- $\varrho$ polytrope), some parameters of the sonic point ($a_\ast^2$, $U_\ast$ and the ratio $m_\ast / R_\ast$) are the same in the full model with backreaction as in the test fluid approximation, provided that the asymptotic parameters are the same in both cases. This shows that for some purposes the knowledge of much simpler test fluid approximation is satisfactory. Although this fact has been proved only for the $\tilde p$ -- $\varrho$ polytropic equation of state, the numerical calculations show that it holds true also for the $\tilde p$ -- $n$ polytropes. We conjecture that it is also valid for a wider subclass of barotropic models.

We will deal with the test fluid approximation until the end of this paper. Among the presented results the special emphasis will be given to the bounds on the speed of sound at the sonic point for a class of barotropic equations of state. If the above conjecture is true, this result will hold also for the general case that includes the backreaction.

%%%%%%%%%%%%%%%%%%%%%%%%%%%%%%%%%%%%%%%%%%%%%%%%%%%%%%%%%%%%%%%%%%%%%%%%%%%%%%%%

\subsubsection{The existence of the solutions}

Within the test fluid approximation the following existence theorem can be proved.

\begin{theorem}
Assume that there exists a sonic point, the backreaction effects can be neglected, the equation of state is a barotrope given by (\ref{barotropic_eos_n}) with the restriction that $1 < \Gamma(n,n_0)$. If in addition the inequality
\begin{equation}
\label{condition_on_eos}
n \frac{d}{dn} \ln a^2 < 2 \left( a^2 + \frac{1}{3} \right)
\end{equation}
holds at the sonic point, then outside the event horizon there exist at least two solutions that bifurcate from the sonic point.
\end{theorem}

Proof. Let us start with the Euler equation (\ref{integrated_euler_eqn}) expressed in the form
\begin{equation}
\label{l_and_p_definition}
L \equiv \frac{\varrho + \tilde p}{n} = \frac{B}{N} = \frac{B}{\sqrt{1 - \frac{2m}{R} + \frac{A^2}{R^4 n^2}}} \equiv P.
\end{equation}
The functions $L$ and $P$ denoting the left and right-hand side of this equation play a crucial role in the entire proof.

From relation (\ref{rho_as_a_function_of_n}) we have
\begin{equation}
\label{left_hand_side}
L = \frac{\tilde p}{n} + 1 + \int_0^n dn^\prime \frac{\tilde p (n^\prime)}{{n^\prime}^2}.
\end{equation}
If the equation of state will be of the form (\ref{barotropic_eos_n}) and the restriction $1 < \Gamma_\mathrm{min} \leqslant \Gamma(n,n_0)$ will hold, then, as it has been already stated,  $\tilde p / n \to 0$ as $n \to 0$. Additionally, the integral appearing in the formula (\ref{left_hand_side}) converges and $\lim_{n \to 0} \int_0^n dn^\prime \tilde p (n^\prime) / {n^\prime}^2 = 0$. Due to these facts we conclude that $L \to 1$ as $n \to 0$ and, what follows simply from the equation (\ref{l_and_p_definition}), $P \to 0$. The other limit, i.e., that of $n \to \infty$ gives, basing on the same assumptions, $L \to \infty$  while for $P$ we have $\lim_{n \to \infty} P = B / \sqrt{1 - 2m/R} < \infty$. Thus in both cases, $n \to 0$ and $n \to \infty$, the same inequality $L > P$ holds independently of the value of $R > 2m$. Now, if it happens that for some value of $n$ (denoted as $n_x$) and a fixed value of $R$ an opposite inequality $L < P$ is satisfied, it follows from the continuity argument that the graphs of $L$ and $P$ must intersect at least in two points, i.e., at least two solutions exist at $R$.

The suitable value of $n_x$ for a given $R$ can be chosen as $n_x = n_\ast \left( R / R_\ast \right)^x$. For the function $n_x(R)$ one can show that at $R = R_\ast$ we have $\partial_R L(n_x) = \partial_R P(n_x)$. Careful computation of the second derivatives gives in turn
\[
\left. \partial_R^2 P(n_x) \right|_{R=R_\ast} = P(R_\ast) \frac{a_\ast^2}{R_\ast^2} \left( \left( 3 a_\ast^2 - 2 \right) x^2 - 9 x - 6 \right)
\]
and
\[\left. \partial_R^2 L(n_x) \right|_{R=R_\ast} = L(R_\ast) \frac{a_\ast^2}{R_\ast^2} \left( \left( a_\ast^2 + \left( n \frac{d}{dn} \ln a^2 \right)_{R=R_\ast} \right) x^2 - x \right).
\]
It is easy to observe that the sign of the expression $\partial_R^2 (P(n_x) - L(n_x))$ taken at the sonic point is equal to the sign of
\[
\left( 2 a_\ast^2 - 2 - \left( n \frac{d}{dn}\ln a^2 \right)_{R = R_\ast} \right) x^2 - 8 x - 6.
\]
The discriminant of the above polynomial of $x$ is always positive provided that inequality (\ref{condition_on_eos}) holds at the sonic point. In this case it is possible to choose such value of $x$ that $\left. \partial_R^2 L (n_x) \right|_{R=R_\ast} < \left. \partial_R^2 P(n_x) \right|_{R=R_\ast}$. This in turn ensures that at least in some vicinity of $R_\ast$ we have $L(n_x) < P(n_x)$.

The domain of the existence of the solution can be extended to the whole domain outside the horizon by exploiting the observation that
\begin{equation}
\label{derivative_of_l-p}
\partial_n (L - P) = \frac{a^2 L}{n} \left( 1 - \frac{4 U^2}{p^2 R^2 a^2} \right).
\end{equation}
The argument allowing us to conclude about this extension is as follows. Assume that there exists $R_1$ (say $R_1 > R$) such that both solutions exist for $R_\ast < R < R_1$ and that there is no intersection of the graphs of $L(n)$ and $P(n)$ for $R > R_1$. Again it follows simply from the continuity that for $R = R_1$ there exist just one point where the two graphs intersect. Obviously, the derivatives with respect to $n$ of both $L$ and $P$ would be equal in this point. That, in turn, contradicts formula (\ref{derivative_of_l-p}) giving a non-zero value for $\partial_n (L - P)$ as long as $R \neq R_\ast$ and as long as we are dealing with a finite baryonic density $n$ and a positive value of the sound speed. Since we have already shown that for $n \to \infty$ we have $L \neq P$ in the region outside the horizon, the baryonic density must remain finite and we are left with the conclusion that at least two solutions exist in the domain described by $R \in (2m, R_\infty)$.

It should be stated that for both polytropic equations of state, i.e., for a type $\tilde p$ -- $\varrho$ and for a type $\tilde p$ -- $n$ polytrope condition (\ref{condition_on_eos}) is satisfied provided that the suitable polytropic exponent is less than 5/3.

%%%%%%%%%%%%%%%%%%%%%%%%%%%%%%%%%%%%%%%%%%%%%%%%%%%%%%%%%%%%%%%%%%%%%%%%%%%%%%%%

\subsubsection{The polytropic equation of state}
\label{the_polytropic_equation_of_state}

The original model of Malec \cite{malec1999} uses the polytropic equation of state of the $\tilde p$ -- $\varrho$ type. We will adhere now the alternative and probably more popular equation of state
\begin{equation}
\label{p_n_polytrope}
\tilde p = K n^\Gamma, \;\; 1 < \Gamma \leqslant 5/3.
\end{equation}

Introducing a new function $\xi$ defined as $\xi = \Gamma  K n^{\Gamma - 1} / (\Gamma - 1)$ one can immediately integrate the Euler equation (\ref{euler_eqn}) to obtain
$N (1 + \xi) = 1 + \xi_\infty$,
where $\xi_\infty$ denotes the value of $\xi$ at the outer boundary of the cloud.

The sound velocity can be also conveniently expressed with the use of $\xi$, $a^2 = (\Gamma - 1) \xi / (1 + \xi)$. It follows that
\begin{equation}
\label{bound_on_a}
a^2 < \Gamma - 1
\end{equation}
is a strict bound for the sound velocity. No such bound exists for the alternative $\tilde p$ -- $\varrho$ polytrope.

The Euler equation integrated above can be also expressed in terms of the sound velocity
\begin{equation}
\label{polytropic_euler_eqn}
\Gamma - 1 - a^2 = N \left( \Gamma - 1 - a_\infty^2 \right).
\end{equation}
Notice that the square of the speed of sound is directly proportional to the lapse function $N$ here. This fact results with the significant simplification of the equations when compared to the $\tilde p$ -- $\varrho$ model.

It has been already shown in \cite{malec1999} that there always exists a unique sonic point for the $\tilde p$ -- $\varrho$ model in which the backreaction is neglected. One can parallel Malec's proof also for the $\tilde p$ -- $n$ case simply by analysing the sonic point equation
\begin{equation}
\Gamma - 1 - a_\ast^2 = N_\ast \left( \Gamma - 1 - a_\infty^2 \right).
\label{sonic_point_eqn}
\end{equation}
It follows that there always exists a unique sonic point in the $\tilde p$ -- $n$ model without the backreaction (i.e., a unique solution to the above equation) provided that $\Gamma \leqslant 5/3$. Moreover, it must be located outside the horizon, i.e., $R_\ast > 2m$.

In this point the $\tilde p$ -- $n$ model differs from the $\tilde p$ -- $\varrho$ model. While in the $\tilde p$ -- $\varrho$ model the sonic point may exist, at least formally, below the horizon\footnote{Such situation could contradict the established views on the properties of matter. If the square of the sound speed in the $\tilde p$ -- $\varrho$ polytropic model exceeds the polytropic index $\Gamma$ we have $\tilde p > \varrho$, i.e., one of the energetic conditions breaks down. This is since for $\tilde p$ -- $\varrho$ polytrope one has
\[ \tilde p = K \varrho^\Gamma = \frac{a^2}{\Gamma} \varrho. \]
}, the $\tilde p$ -- $n$ equation of state does not allow (even theoretically) such a possibility to happen.

Further progress can be made by considering equation (\ref{sonic_point_eqn}) squared and rewritten in the form
\begin{eqnarray}
\lefteqn{3 \left( a_\ast^2 \right)^3 + (7 - 6\Gamma) \left( a_\ast^2 \right)^2 + (\Gamma - 1)(3 \Gamma - 5) a_\ast^2 +} \nonumber \\
&& + a_\infty^2 \left( 2(\Gamma - 1) - a_\infty^2 \right) = 0.
\label{squared_sonic_point_eqn}
\end{eqnarray}

This is a purely algebraic cubic equation for $a_\ast^2$. It is easy to show that for $\Gamma \leqslant 5/3$ and for the interesting range of $a_\infty^2$, i.e., $0 < a_\infty^2 < \Gamma - 1$ three real different roots of this equation exist. Since the original unsquared equation (\ref{sonic_point_eqn}) can be proved to have a unique solution, only one of the mentioned roots of (\ref{squared_sonic_point_eqn}) is the one we search for. Careful analysis shows that the first of two remaining solutions lays outside the range $0 < a_\ast^2 < 1$ and the second simply does not satisfy the original equation (\ref{sonic_point_eqn}). The only physical solution can be expressed by the following Cardano's formula (see also \cite{das})
\begin{eqnarray}
a_\ast^2 & = & \frac{1}{9} \biggr\{ 6\Gamma - 7 + 2(3\Gamma - 2) \cos \biggm[ \frac{\pi}{3} + \frac{1}{3} \arccos \biggl\{ \frac{1}{2(3\Gamma - 2)^3} \Big( 54 \Gamma^3 + \nonumber \\
& & - 351 \Gamma^2  - 558 \Gamma + 486 (\Gamma - 1) a_\infty^2 - 243 a_\infty^4 - 259 \Big) \biggl\} \biggm] \biggr\}.
\label{solution_for_a}
\end{eqnarray}
One may check, that for $a_\infty^2 = \Gamma - 1$ the above gives $a_\ast^2 = a_\infty^2$ and for $a_\infty^2 = 0$ we get $a_\ast^2 = 0$.

The formula (\ref{integrated_U_eqn}) for the value of the velocity $U$ at a given radius $R$ can be transformed into the form involving the sound speed instead of the baryonic density. For the $\tilde p$ -- $n$ polytrope we have
\[ U = U_\ast \frac{R_\ast^2}{R^2} \left( \frac{a_\ast^2}{a^2} \frac{\Gamma - 1 - a^2}{\Gamma - 1 - a_\ast^2} \right)^\frac{1}{\Gamma - 1}. \]
This equation, together with the Euler equation (\ref{polytropic_euler_eqn}) and equations (\ref{integrated_lapse_eqn}) and (\ref{hamiltonian_constraint}), constitutes the complete set describing the full polytropic model that takes into account the backreaction effects. Since in the test fluid approximation we have obtained an analytic expression for the sound speed in the sonic point and
\begin{equation}
\label{u_n_r_ast}
U_\ast^2 = \frac{a_\ast^2}{1 + 3 a_\ast^2}, \;\;\; N_\ast = \frac{1}{\sqrt{1 + 3 a_\ast^2}}, \;\;\; R_\ast = \frac{m}{2} \frac{1 + 3 a_\ast^2}{a_\ast^2},
\end{equation}
i.e., we know precisely all quantities characterising the sonic point, it is possible to express the accretion rate with just one analytic formula depending on asymptotic parameters $a_\infty$ and $n_\infty$. Indeed, after some calculations one can arrive at the form
\begin{eqnarray}
\dot m & = & - 4 \pi N R^2 U (\varrho + \tilde p) = \nonumber \\
& = & \pi n_\infty m^2 \frac{\Gamma - 1}{\Gamma - 1 -a_\infty^2} \left( \frac{1 + 3a_\ast^2}{a_\ast^2} \right)^\frac{3}{2} \left( \frac{a_\ast^2}{a_\infty^2} \frac{\Gamma - 1 - a_\infty^2}{\Gamma - 1 - a_\ast^2} \right)^\frac{1}{\Gamma - 1}.
\label{polytropic_accretion_rate}
\end{eqnarray}
In this point the model with the $\tilde p$ -- $n$ polytrope appears to be simpler than the alternative $\tilde p$ -- $\varrho$ model in which the value of the accretion rate can be only estimated (or computed numerically) even when the backreaction is neglected. The detailed comparison of the two models can be found in \cite{kinasiewicz_lanczewski}.

Let us also notice that some authors (see e.g. \cite{shapiro_teukolsky}) define an accretion rate as the time derivative of the baryonic mass (instead of the quasilocal one). The accretion rate defined this way can be expressed by the formula $\dot m_\mathrm{bar} = (\partial_t - (\partial_t R)\partial_R) m_\mathrm{bar}(R) = - 8 \pi N U R n / p$. In the test fluid approximation where $N \approx pR/2$ we get $\dot m_\mathrm{bar} = - 4 \pi R^2 U n$.
Finally, equation (\ref{integrated_euler_eqn}) allows one to show the following relation between the two rates $\dot m = B \dot m_\mathrm{bar} = \left( (\varrho_\infty + \tilde p_\infty) / n_\infty \right) \dot m_\mathrm{bar}$. Both these accretion rates become equal for $\varrho_\infty = n_\infty$, $\tilde p_\infty \ll \varrho_\infty$, i.e., when the fluid at the outer boundary is nonrelativistic.

%%%%%%%%%%%%%%%%%%%%%%%%%%%%%%%%%%%%%%%%%%%%%%%%%%%%%%%%%%%%%%%%%%%%%%%%%%%%%%%%

\subsubsection{Estimates onto the speed of sound}
\label{sound_speed_at_the_sonic_point_for_the_barotropic_model}

The analytic solution of the sonic point equation for the $\tilde p$ -- $n$ polytropic model in the test fluid approximation allows one to make estimates onto the sound speed at the sonic point for a wider class of barotropes, namely for those with monotonic barotropic exponents.

Indeed, let us assume a barotropic equation of state expressed as
\begin{equation}
\label{barotropic_eos_for_estimates}
\tilde p (n) = \tilde p (n_-) \left( \frac{n}{n_-} \right)^{\Gamma(n,n_-)} = \tilde p (n_+) \left( \frac{n}{n_+} \right)^{\Gamma(n,n_+)},
\end{equation}
where densities of reference $n_-$ and $n_+$ are chosen in a way which ensures that $n_- \leqslant n$ and $n_+ \geqslant n$  (i.e., $n_-$ should be less or equal to the smallest baryonic density appearing in the model\footnote{This means, in practice, $n_- \leqslant n_\infty$} while $n_+$ should be greater than the largest baryonic density\footnote{For most cases it should suffice to have $n_+ \geqslant n(2m)$}). Let us additionally assume that $1 < \Gamma_\mathrm{min}^- \leqslant \Gamma(n,n_-) \leqslant \Gamma_\mathrm{max}^-$ and $1 < \Gamma_\mathrm{min}^+ \leqslant \Gamma(n,n_+) \leqslant \Gamma_\mathrm{max}^+$ for some constants $\Gamma_\mathrm{min}^-$, $\Gamma_\mathrm{max}^-$, $\Gamma_\mathrm{min}^+$ and $\Gamma_\mathrm{max}^+$.

\begin{lemma}
\label{th_bound_for_nonincreasing_gamma}
Consider a spherically symmetric, steady accretion onto the black hole with a negligible backreaction. Let the equation of state be expressed in the form (\ref{barotropic_eos_for_estimates}) and let both exponents $\Gamma(n,n_-)$ and $\Gamma(n,n_+)$ be nonincreasing functions of the baryonic density $n$. Then the square of the sound speed at the sonic point is bounded by two analytic expressions
$ a_-^2 \leqslant a_\ast^2 \leqslant a_+^2 $,
where
\begin{eqnarray}
a^2_- & = & \frac{1}{9} \Biggr\{ 6\Gamma^+_\mathrm{min} - 7 + 2 \left( 3\Gamma^+_\mathrm{min} - 2 \right) \cos \Biggm[ \frac{\pi}{3} + \\
& & + \frac{1}{3} \arccos \Biggl\{ \frac{1}{2 \left( 3\Gamma^+_\mathrm{min} - 2 \right)^3} \Biggl( 2 \left( 27 \left( \Gamma^+_\mathrm{min} \right)^3 -  54 \left( \Gamma^+_\mathrm{min} \right)^2 + \right. \nonumber \\
& & + \left. 36 \Gamma^+_\mathrm{min} - 8 \right) - 243 \left( \Gamma^+_\mathrm{min} - 1 \right)^2 \left( \frac{\Gamma^-_\mathrm{max} - 1 - a^2_\infty}{\Gamma^-_\mathrm{max} - 1} \right)^2 \Biggl) \Biggl\} \Biggm] \Biggr\} \nonumber
\label{oszacowanie_od_dolu}
\end{eqnarray}
and
\begin{eqnarray}
a^2_+ & = & \frac{1}{9} \Biggr\{ 6\Gamma^-_\mathrm{max} - 7 + 2 \left( 3\Gamma^-_\mathrm{max} - 2 \right) \cos \Biggm[ \frac{\pi}{3} + \\
& & + \frac{1}{3} \arccos \Biggl\{ \frac{1}{2 \left( 3\Gamma^-_\mathrm{max} - 2 \right)^3} \Biggl( 2 \left( 27 \left( \Gamma^-_\mathrm{max} \right)^3 -  54 \left( \Gamma^-_\mathrm{max} \right)^2 + \right. \nonumber \\
& & + \left. 36 \Gamma^-_\mathrm{max} - 8 \right) - 243 \left( \Gamma^-_\mathrm{max} - 1 \right)^2 \left( \frac{\Gamma^+_\mathrm{min} - 1 - a^2_\infty}{\Gamma^+_\mathrm{min} - 1} \right)^2 \Biggl) \Biggl\} \Biggm] \Biggr\}. \nonumber
\label{oszacowanie_od_gory}
\end{eqnarray}
\end{lemma}

A completely analogous lemma can be formulated for nondecreasing barotropic exponents. 

\begin{lemma}
\label{th_bound_for_nondecreasing_gamma}
Let the conditions of lemma \ref{th_bound_for_nonincreasing_gamma} be satisfied with an exception that this time both exponents $\Gamma(n,n_-)$ and $\Gamma(n,n_+)$ are assumed to be nondecreasing functions of $n$. Then the square of the sound speed computed at the sonic point is bounded from above and from below by analogous expressions
$ a^2_- \leqslant a^2_\ast \leqslant a^2_+ $,
where
\begin{eqnarray*}
a^2_- & = & \frac{1}{9} \Biggr\{ 6\Gamma^-_\mathrm{min} - 7 + 2 \left( 3\Gamma^-_\mathrm{min} - 2 \right) \cos \Biggm[ \frac{\pi}{3} + \nonumber \\
& & + \frac{1}{3} \arccos \Biggl\{ \frac{1}{2 \left( 3\Gamma^-_\mathrm{min} - 2 \right)^3} \Biggl( 2 \left( 27 \left( \Gamma^-_\mathrm{min} \right)^3 -  54 \left( \Gamma^-_\mathrm{min} \right)^2 + \right. \nonumber \\
& & + \left. 36 \Gamma^-_\mathrm{min} - 8 \right) - 243 \left(\Gamma^-_\mathrm{min} - 1 \right)^2 \left( \frac{\Gamma^+_\mathrm{max} - 1 - a^2_\infty}{\Gamma^+_\mathrm{max} - 1} \right)^2\Biggl) \Biggl\} \Biggm] \Biggr\}
\end{eqnarray*}
and
\begin{eqnarray*}
a^2_+ & = & \frac{1}{9} \Biggr\{ 6\Gamma^+_\mathrm{max} - 7 + 2 \left( 3\Gamma^+_\mathrm{max} - 2 \right) \cos \Biggm[ \frac{\pi}{3} + \nonumber \\
& & + \frac{1}{3} \arccos \Biggl\{ \frac{1}{2 \left( 3\Gamma^+_\mathrm{max} - 2 \right)^3} \Biggl( 2 \left( 27 \left( \Gamma^+_\mathrm{max} \right)^3 -  54 \left( \Gamma^+_\mathrm{max} \right)^2 + \right. \nonumber \\
& & + \left( 36 \Gamma^+_\mathrm{max} - 8 \right) - 243 \left( \Gamma^+_\mathrm{max} - 1 \right)^2 \left( \frac{\Gamma^-_\mathrm{min} - 1 - a^2_\infty}{\Gamma^-_\mathrm{min} - 1} \right)^2 \Biggl) \Biggl\} \Biggm] \Biggr\}.
\end{eqnarray*}
\end{lemma}

The proofs of these lemmas have been omitted since they are straightforward.

%%%%%%%%%%%%%%%%%%%%%%%%%%%%%%%%%%%%%%%%%%%%%%%%%%%%%%%%%%%%%%%%%%%%%%%%%%%%%%%%

\subsubsection{Bounds on the accretion rate in the barotropic case}
\label{bounds_on_the_accretion_rate}

Having estimated the sound speed at the sonic point we should be able to obtain bounds on the accretion rate $\dot m$ as well. We will restrict ourselves to the case of lemma \ref{th_bound_for_nonincreasing_gamma}. The bounds on the accretion rate for the alternative case described in statement \ref{th_bound_for_nondecreasing_gamma} can be obtained in an analogous way.

Equations (\ref{u_n_r_ast}) (valid in the test fluid case) give the values of $U_\ast$, $N_\ast$ and $R_\ast$ expressed in terms of the sound speed $a_\ast$. Thus, as it is clear from formula (\ref{accretion_rate}), the only quantity that needs to be estimated is $\varrho_\ast + \tilde p_\ast$.

Suppose now that the conditions of statement \ref{th_bound_for_nonincreasing_gamma} hold. One can find $ \Gamma(n,n_+) \tilde p / a^2 \leqslant \varrho + \tilde p \leqslant \Gamma(n,n_-) \tilde p / a^2$. Furthermore, making use of the same inequalities together with their asymptotic versions (i.e., at $R = R_\infty$) we can write
\[ \tilde p \leqslant \tilde p(n_0) \left( \left( \frac{n_\infty}{n_0} \right)^{\Gamma(n_\infty,n_0)-1} \frac{a^2}{a_\infty^2} \frac{\Gamma(n_\infty,n_-)}{N \Gamma(n,n_+)}  \right)^\frac{\Gamma(n,n_0)}{\Gamma(n,n_0) - 1} \]
and similarly
\[ \tilde p \geqslant \tilde p(n_0) \left( \left( \frac{n_\infty}{n_0} \right)^{\Gamma(n_\infty,n_0)-1} \frac{a^2}{a_\infty^2} \frac{\Gamma(n_\infty,n_+)}{N \Gamma(n,n_-)}  \right)^\frac{\Gamma(n,n_0)}{\Gamma(n,n_0) - 1} \] 
where $n_0$ stands for $n_+$ or $n_-$.

It follows that the bounds on the accretion rate $\dot m$ can be formulated as
\begin{eqnarray*}
\dot m & = & \pi m^2 \frac{1 + 3 a_\ast^2}{a_\ast^3} \left( \varrho_\ast + \tilde p_\ast \right) \leqslant \\
& \leqslant & \pi m^2 \frac{1 + 3 a_\ast^2}{a_\ast^3} \frac{\Gamma^-_\mathrm{max}}{a_\ast^2} \tilde p(n_-) \left( \frac{n_\infty}{n_-} \right)^{\Gamma^-_\mathrm{max}} \left( \frac{a_\ast^2 \sqrt{1+3 a_\ast^2}}{a_\infty^2} \frac{\Gamma^-_\mathrm{max}}{\Gamma^+_\mathrm{min}}  \right)^\frac{\Gamma^-_\mathrm{max}}{\Gamma^-_\mathrm{max} - 1}
\end{eqnarray*}
and
\begin{eqnarray*}
\dot m & \geqslant & \pi m^2 \frac{1 + 3 a_\ast^2}{a_\ast^3} \frac{\Gamma^+_\mathrm{min}}{a_\ast^2} \tilde p(n_+) \left( \frac{n_\infty}{n_+} \right)^{\Gamma^+_\mathrm{max}} \times \\
&& \times \left( \frac{\Gamma^+_\mathrm{min}}{\Gamma^-_\mathrm{max}} \right)^\frac{\Gamma^+_\mathrm{max}}{\Gamma^+_\mathrm{max}-1} \left( \frac{a_\ast^2 \sqrt{1 + 3 a_\ast^2}}{a_\infty^2} \right)^\frac{\Gamma^+_\mathrm{min}}{\Gamma^+_\mathrm{min} - 1}.
\end{eqnarray*}
Obviously, the above formulae together with formulae (\ref{oszacowanie_od_dolu}) and (\ref{oszacowanie_od_gory}) giving the lower and upper bounds for the value of $a_\ast^2$ complete our task. One can also check that these inequalities are exact. They are saturated for the $\tilde p$ -- $n$ polytropes in which case their right-hand sides reduce to the same expression (\ref{polytropic_accretion_rate}).

%%%%%%%%%%%%%%%%%%%%%%%%%%%%%%%%%%%%%%%%%%%%%%%%%%%%%%%%%%%%%%%%%%%%%%%%%%%%%%%%
%%%%%%%%%%%%%%%%%%%%%%%%%%%%%%%%%%%%%%%%%%%%%%%%%%%%%%%%%%%%%%%%%%%%%%%%%%%%%%%%

\section{Closing remarks}
\label{closing_remarks}

The stability of the self-gravitating polytropes against radial perturbations has been examined by Tooper \cite{tooper}. All configurations with the negative binding energy happen to be unstable, however the positivity of the binding energy is only a necessary condition of the stability of the configuration. The problem of the stability of the self-gravitating configurations of general barotropic fluids remains (to our knowledge) still open; it is, however, intuitively tempting to treat the sign of the binding energy of such configuration as carrying information about the stability. Thus inequalities presented here, relating the maximal speed of sound within the considered configuration and the sign of its binding energy, might be treated as some versions of the Jeans inequality. These inequalities, being straightforward generalisations of those given by Karkowski and Malec constitute the necessary conditions for the negativity of the binding energy. Generalisation of the sufficient condition formulated in \cite{KM} remains still an open task.

As to the steady spherically symmetric accretion we believe that at least from the relativist's point of view the issue of outermost interest is the backreaction of the fluid onto the space-time metric. This effect has been investigated so far only for the polytropic case \cite{kkmms}. In this paper we were mostly dealing with the test fluid approximation obtaining information about the characteristics of the sonic point for a class of barotropic equations of state. Since the characteristics of the sonic point are expected to be the same in the full picture that includes the backreaction, our results should be relevant in the general, spherically symmetric case.

One of the most striking effects caused by the backreaction is that the accretion rate achieves a maximum value for the ratio of the mass of the fluid versus the total mass of the system being equal to 1/3. This result has been obtained both for the $\tilde p$ -- $\varrho$ and $\tilde p$ -- $n$ polytrope. We conjecture it holds true also for a wider class of barotropic equations of state.

\section*{Acknowledgements}

We wish to express our thanks to Professor Edward Malec for his help and many fruitful discussions.

%%%%%%%%%%%%%%%%%%%%%%%%%%%%%%%%%%%%%%%%%%%%%%%%%%%%%%%%%%%%%%%%%%%%%%%%%%%%%%%

\end{document}